\documentclass[10pt,twocolumn,twoside,journal]{IEEEtran}
\usepackage{subfigure}
\usepackage{setspace}
\usepackage{amsmath}
\usepackage{amssymb}
\usepackage{amsfonts}
\usepackage{amscd}
\usepackage{mathrsfs}
\usepackage[final]{graphicx}
\usepackage{graphicx}
\usepackage{psfrag}
\usepackage{url}
\usepackage{textcomp}
\usepackage{multirow}

\newcommand{\Define}{\stackrel{\triangle}{=}}

\begin{document}

\markboth{}{N. Srinidhi \MakeLowercase{\textit{et al.}}: 
Near-ML Signal Detection in Large-Dimension Linear Vector
Channels Using Reactive Tabu Search}

\title{{\Huge Near-ML Signal Detection in Large-Dimension \\ Linear Vector
Channels Using Reactive Tabu Search}
}
\author{N. Srinidhi, Saif K. Mohammed, A. Chockalingam, 
and B. Sundar Rajan 
\thanks{This work in part was presented in IEEE ISIT'2009, Seoul, Korea,
July 2009, and is accepted for presentation in IEEE GLOBECOM'2009, 
Honolulu, USA, December 2009. \newline
The authors are with the Department of Electrical Communication 
Engineering, Indian Institute of Science, Bangalore-560012, India.
}}

\markboth{N. Srinidhi \MakeLowercase{\textit{et al.}}:
Near-ML Signal Detection in Large-Dimension Linear Vector
Channels Using Reactive Tabu Search}
{N. Srinidhi \MakeLowercase{\textit{et al.}}: 
Near-ML Signal Detection in Large-Dimension Linear Vector
Channels Using Reactive Tabu Search}

\maketitle

\begin{abstract}
Low-complexity near-optimal signal detection in large dimensional
communication systems is a challenge. In this paper, we present a
reactive tabu search (RTS) algorithm, a heuristic based combinatorial
optimization technique, to achieve low-complexity near-maximum
likelihood (ML) signal detection in linear vector channels with large
dimensions. Two practically important large-dimension linear vector
channels are considered: $i)$ multiple-input multiple-output (MIMO)
channels with large number (tens) of transmit and receive antennas,
and $ii)$ severely delay-spread MIMO inter-symbol interference (ISI)
channels with large number (tens to hundreds) of multipath components.
These channels are of interest because the former offers the benefit
of increased spectral efficiency (several tens of bps/Hz) and the
latter offers the benefit of high time-diversity orders. Our simulation
results show that, while algorithms including variants of sphere decoding
do not scale well for large dimensions, the proposed RTS algorithm scales
well for signal detection in large dimensions while achieving increasingly
closer to ML performance for increasing number of dimensions.
\end{abstract}


\vspace{-2mm}
\section{Introduction}
\label{sec1}
\vspace{-1mm}
Large-dimension communication systems are likely to play an important
role in modern wireless communications, where dimensions can be in space,
time, frequency and their combinations. Large dimensions can bring several
advantages with respect to the performance of communication systems. For
example, use of large number of transmit/receive antennas increases the
number of spatial dimensions, which results in increased capacity
\cite{telatar},\cite{paul}. A severely delay-spread inter-symbol
interference (ISI) channel (i.e., large number of echoes of the
transmitted signal in time dimension), as witnessed in ultrawideband
(UWB) systems, can provide the opportunity for increased time-diversity
\cite{ngoc}. Harnessing such benefits of large-dimensions
in practice, however, is challenging. In particular, optimum
receiver complexity can become practically infeasible in large dimensions.
Consequently, low-complexity receiver techniques/algorithms that scale
well for large dimensions while achieving near-optimal performance are
of interest. It has been found that many modern meta-heuristic algorithms
give near-optimal performance at a much reduced complexity \cite{reeves}.
In this paper, we report one such heuristic based on {\em tabu search}
\cite{tabu1},\cite{tabu2}, and illustrate its near-optimal
performance in two practically important large dimension systems, namely
$i)$ a `large-MIMO system' with {\em tens of transmit/receive antennas}
(with a motivation to achieve high spectral efficiencies), and $ii)$ a
severely delay-spread MIMO UWB system with {\em tens to hundreds of
multipath components} (with a motivation to achieve high time-diversity
orders).

Tabu search (TS) is a heuristic originally designed to obtain approximate
solutions to combinatorial optimization problems \cite{tabu1}-\cite{tabu3}.
TS is increasingly being applied in communication problems
\cite{tabu4}-\cite{tabu6}. For e.g., in \cite{tabu4}, design of
constellation label maps to maximize asymptotic coding gain is formulated
as a quadratic assignment problem, which is solved using a reactive TS
(RTS) strategy \cite{tabu3}. RTS approach is shown to be effective in
terms of bit error performance and efficient in terms of computational
complexity in CDMA multiuser detection \cite{tabu5}. In \cite{tabu6},
a {\em fixed} TS based detection in V-BLAST is presented for small number
of antennas. A key objective in this paper is to propose a {\em reactive}
tabu search based approach to seek approximately maximum-likelihood (ML)
solutions in large dimension problems (but with significantly lower
computational complexity than that of the true ML solution) in linear
vector channels (LVC) in general, and to establish its performance and
complexity in two interesting communication systems in particular.

The first communication system we consider is a large-MIMO system that
employs tens of transmit antennas to achieve high spectral efficiencies
-- e.g., a $16\times 16$ V-BLAST system with 16-QAM and rate-3/4 turbo
code can achieve a spectral efficiency of 48 bps/Hz. We show that the
RTS algorithm achieves increasingly closer to ML performance for
increasing number of transmit antennas (we refer to this behavior of
the algorithm as the `large-dimension behavior'). For e.g., in a
$64\times 64$ V-BLAST system with 4-QAM, RTS is shown to achieve $10^{-3}$
uncoded BER at an SNR of just 0.4 dB away from single-input single-output
(SISO) AWGN performance. We present a comparison of the performance and
complexity of RTS with those of low-complexity variants of sphere decoders
(SD), including a suboptimal fixed-complexity SD (FSD) reported in \cite{fsd}.
In a $32\times 32$ V-BLAST system with 4-QAM, RTS is shown to perform better
than FSD by about 1.5 dB at $10^{-2}$ uncoded BER. Interestingly, RTS
achieves this better performance at about an order less complexity than
FSD. We also show that RTS can achieve near-ML performance in decoding
large non-orthogonal space-time codes (STBCs) from cyclic division
algebras (CDA), which can offer full transmit diversity in addition to
achieving full rate as in V-BLAST \cite{bsr},\cite{cda}.

The second communication scenario considered is equalization in severely
delay-spread MIMO-ISI UWB channels with large number of multipath
components (MPC). Communication systems using UWB techniques typically
have very high transmission bandwidths to accommodate very high data
rates \cite{ngoc}. Such UWB channels are characterized by severe ISI
due to large delay spreads \cite{uwb0}-\cite{uwb3}. The number of MPCs
in indoor and industrial environments has been observed to be of the
order of several tens to hundreds; number of MPCs ranging from 12 to
120 are common in UWB channel models \cite{uwb0},\cite{uwb3}. These MPCs,
if carefully exploited, can provide the opportunity to achieve increased
time-diversity benefits \cite{uwb0}. Algorithms based on likelihood 
ascent search (LAS)/bit flipping \cite{jsac},\cite{jstsp},\cite{akino} 
and factor graphs \cite{wo} have been proposed for equalization in such
systems. We show that the proposed RTS algorithm achieves increasingly 
close to optimal performance for increasing number of MPCs, and achieves
better performance due its inherent escape strategy from local minima.

The rest of the paper is organized as follows. The proposed RTS algorithm
for detection in linear vector channels is presented in Section \ref{sec2}.
BER performance and complexity of the RTS algorithm in comparison with
those of other detectors including variants of sphere decoders are
presented in Sections \ref{sec3} to \ref{sec5}. Conclusions are presented
in Sections \ref{sec6}.

\vspace{-1mm}
\section{Proposed RTS Based Detection in LVCs}
\label{sec2}
We consider linear vector channels where
a $d_t$-dimensional input vector\footnote{Notation: Vectors and matrices
are denoted by boldface lowercase letters and boldface uppercase letters,
respectively. $(.)^*$, $[.]^T$, and $[.]^H$ denote conjugation,
transpose and Hermitian operations, respectively. $|.|$ denotes the
absolute value operator. ${\bf A}(i,j)$ denotes
the element in the $i$th row and $j$th column of matrix ${\bf A}$. $a_i$
denotes the $i$th element of the vector ${\bf a}$. $\Re(.)$ and $\Im(.)$
denote the real and imaginary parts of a complex argument, and
${\mathfrak j}=\sqrt{-1}$. ${\bf I}_n$ denotes the $n\times n$
identity matrix.} ${\bf x} \in {\mathbb A}^{d_t}$ (${\mathbb A}$
denotes a finite set from the complex field) is linearly transformed by a
$d_r\times d_t$ channel transfer matrix,
${\bf H} \in {\mathbb C}^{d_r\times d_t}$, and is corrupted by a
$d_r$-dimensional noise vector, ${\bf n} \in {\mathbb C}^{d_r}$, so that
the $d_r$-dimensional output vector, ${\bf y} \in {\mathbb C}^{d_r}$, is
given by
\begin{eqnarray}
{\bf y} & = & {\bf H}{\bf x} + {\bf n}.
\label{eqn1}
\end{eqnarray}
In communication systems, ${\bf x}$ and ${\bf y}$ can be the transmitted
and received signal vectors, respectively, and the goal is to obtain an
estimate of the transmitted vector ${\bf x}$, given ${\bf y}$ and the
knowledge of ${\bf H}$. When the noise is Gaussian, the maximum-likelihood
(ML) detection rule is given by
\begin{eqnarray}
\hspace{-5mm}
{\widehat {\bf x}}_{ML} & = & {\mbox{arg min}\atop{{\bf x} \in {\mathbb A}^{d_t}}}
\,\,\,
\| {\bf y} - {\bf H}{\bf x}\|^2
\,\,\, =  \,\,\, {\mbox{arg min}\atop{{\bf x} \in {\mathbb A}^{d_t}}}
\,\,\, \phi({\bf x}),
\label{MLdetection}
\end{eqnarray}
where
$\phi({\bf x}) \Define {\bf x}^H {\bf H}^H{\bf H}{\bf x}-2\Re\left({\bf y}^H{\bf H}{\bf x}\right)$.
The computational complexity in (\ref{MLdetection}) is exponential in $d_t$,
which is prohibitive for large $d_t$. Our interest is to achieve near-ML
performance for large $d_t$ at low complexities. In the following subsection,
we present a RTS based detection algorithm which is a low-complexity
iterative local search algorithm suited well for large $d_t$.

\vspace{-2mm}
\subsection{RTS Algorithm}
\label{sec21}
The RTS algorithm starts with an initial solution vector, defines
a neighborhood around it (i.e., defines a set of neighboring vectors
based on a neighborhood criteria),
and moves to the best vector among the neighboring vectors (even if the
best neighboring vector is worse, in terms of likelihood, than the
current solution vector; this allows the algorithm to escape from local
minima). This process is continued for a certain number of iterations,
after which the algorithm is terminated and the best among the
solution vectors in all the iterations is declared as the final solution
vector. In defining the neighborhood of the solution vector in a given
iteration, the algorithm attempts to avoid cycling by making the moves to
solution vectors of the past few iterations as `tabu' (i.e., prohibits these
moves), which ensures efficient search of the solution space. The number
of these past iterations is parametrized as the `tabu period.' The search
is referred to as fixed tabu search if the tabu period is kept constant.
If the tabu period is dynamically changed (e.g., increase the tabu period
if more repetitions of the solution vectors are observed in the search
path), then the search is called reactive tabu search. We consider reactive
tabu search in this paper because of its robustness (choice of a good fixed
tabu period can be tedious).

\vspace{2mm}
{\bfseries {\em Neighborhood Definition:}}
Let $M$ denote the cardinality of ${\mathbb A} = \{a_1,a_2,\cdots,a_M\}$.
Define a set ${\cal N}(a_q)$, $q\in \{1,\cdots,M$\}, as
a fixed subset of ${\mathbb A} \backslash a_q$, which we refer to as the
{\em symbol-neighborhood} of $a_q$. We choose the cardinality of this set
to be the same for all $a_q, \, q=1,\cdots,M$; i.e., we take
$|{\cal N}(a_q)| = N, \, \forall q$. Note that the maximum and minimum
values of $N$ are $M-1$ and 1, respectively. We choose the symbol
neighborhood based on Euclidean distance, i.e., for a given symbol,
those $N$ symbols which are the nearest will form its neighborhood;
the nearest symbol will be the first neighbor, the next nearest symbol
will be the second neighbor, and so on. For e.g.,
{\small ${\mathbb A}=\{-3,-1,1,3\}$} for 4-PAM, and choosing
$N$ to be 2, {\small ${\cal N}(-3)=\{-1,1\}$},
{\small ${\cal N}(-1)=\{-3,1\}$}, {\small ${\cal N}(1)=\{-1,3\}$},
{\small ${\cal N}(3)=\{1,-1\}$} are possible symbol-neighborhoods.
Let $w_v(a_q)$, $v=1,\cdots,N$ denote
the $v$th element in ${\cal N}(a_q)$; i.e., we say $w_v(a_q)$ is the
$v$th symbol-neighbor of $a_q$.

Let
${\bf x}^{(m)}=\small{[x_1^{(m)} \thinspace x_2^{(m)} \cdots x_{d_t}^{(m)}]}$
denote the data vector belonging to the solution space in the $m$th
iteration, where $x_i^{(m)} \in {\mathbb A}$. We refer to the vector
\begin{eqnarray}
\mathbf{z}^{(m)}(u,v) & = & \big[z^{(m)}_1(u,v)
\,\,\,\, z^{(m)}_2(u,v) \, \cdots \,
z^{(m)}_{d_t}(u,v) \big],
\end{eqnarray}
as the $(u,v)$th {\em vector-neighbor} $\big($or simply the $(u,v)$th
neighbor$\big)$
of $\mathbf{x}^{(m)}$,
{\small $u=1,\cdots,d_t$, $v=1,\cdots,N$}, if $i)$ $\mathbf{x}^{(m)}$
differs from $\mathbf{z}^{(m)}(u,v)$ in the $u$th coordinate only, and $ii)$
the $u$th element of $\mathbf{z}^{(m)}(u,v)$ is the $v$th symbol-neighbor
of $x_u^{(m)}$. That is,
\begin{equation}
z_i^{(m)}(u,v) \,\,= \,\, \left\{
\begin{array}{ll}
x^{(m)}_i & \mbox{for} \,\,\, i\neq u \\
w_v(x_u^{(m)}) & \mbox{for} \,\,\, i=u.
\end{array}\right.
\label{eq11}
\end{equation}
So we will have $d_tN$ vectors which differ from a given vector in the
solution space in only one coordinate. These $d_tN$ vectors form the
neighborhood of the given vector. We note that neighborhood definition
based on {\em bit-flipping} \cite{sun},\cite{jsac} is a special case of
the above
neighborhood definition for $M=2$, $N=1$. An operation on ${\bf x}^{(m)}$
which gives ${\bf x}^{(m+1)}$ belonging to the vector-neighborhood of
${\bf x}^{(m)}$ is called a {\em move}. The algorithm is said to
execute a move $(u,v)$ if $\mathbf{x}^{(m+1)} = \mathbf{z}^{(m)}(u,v)$.
We note that the number of candidates to be considered for a move in
any one iteration is $d_tN$. Also, the overall number of `distinct' moves
possible is $d_tMN$, which is the cardinality of the union of all moves
from all $M^{d_t}$ possible solution vectors. The tabu value of a move,
which is a non-negative integer, means that the move cannot be considered
for that many number of subsequent iterations, unless certain conditions
are satisfied.

\vspace{2mm}
{\bfseries {\em Tabu Matrix:}}
A $\textit{tabu\_matrix}$ ${\bf T}$ of size $d_tM\times N$ is the matrix
whose entries denote the tabu values of moves.
For each coordinate of the solution vector (there are $d_t$ coordinates),
there are $M$ rows in ${\bf T}$, where each row corresponds to one
symbol in the modulation alphabet ${\mathbb A}$; the indices of the rows
for the $u$th coordinate are from $(u-1)M+1$ to $uM$, $u\in \{1,\cdots,d_t\}$.
The $N$ columns of ${\bf T}$ correspond to the $N$ symbol-neighbors of
the symbol corresponding to each row.
In other words, the $(r,s)$th entry of the
$\textit{tabu\_matrix}$, $r=1,\cdots,d_tM$, $s=1,\cdots,N$, corresponds
to the move $(u,v)$ from
$\mathbf{x}^{(m)}$ when $u=\lfloor \frac{r-1}{M} \rfloor+1$, $v=s$
and $x_u^{(m)}=a_q$, where $q=mod(r-1,M)+1$.
The entries of the tabu matrix, which are non-negative integers, are updated
in each iteration, and they are used to decide the direction in which the
search proceeds (as described in the algorithm description below).

\vspace{2mm}
{\bfseries {\em Algorithm:}}
Let $\mathbf{g}^{(m)}$ be the vector which has the least ML cost found
till the $m$th iteration of the algorithm. Let $l_{rep}$ be the average
length (in number of iterations) between two successive occurrences of
a solution vector (repetitions). Tabu period, $P$, a dynamic non-negative
integer parameter, is defined as follows:
if a move is marked as tabu in an iteration, it will remain as tabu
for $P$ subsequent iterations unless the move results in a better
solution. A binary flag, $lflag \in \{0,1\}$,
is used to indicate whether the algorithm has reached a local minima
in a given iteration or not; this flag is used in the evaluation of the
stopping criterion of the algorithm. The algorithm starts with an initial
solution vector  ${\bf x}^{(0)}$, which, for e.g., could be the MMSE or
matched filter output vector. Set $\mathbf{g}^{(0)} = {\bf x}^{(0)}$,
$l_{rep}=0$,
and $P=P_0$. All the entries of the $\textit{tabu\_matrix}$ are set to
zero. Define $\thinspace \mathbf{y}_{MF} \Define \mathbf{H}^H\mathbf{y}$,
and $\thinspace\mathbf{R} \Define \mathbf{H}^H\mathbf{H}$. Compute
$\mathbf{y}_{MF}$ and $\mathbf{R}$. The following steps 1) to 3) are
performed in each iteration. Consider $m$th iteration in the algorithm,
$m\geq 0$.

\vspace{2mm}
{\em Step 1):}
Initialize $lflag=0$. Define
$\thinspace\mathbf{f}^{(m)}\Define \mathbf{R}\mathbf{x}^{(m)}-\mathbf{y}_{MF}$.
Let $\mathbf{e}=\mathbf{z}^{(m)}(u,v)-\mathbf{x}^{(m)}$.
The ML costs of the $d_tN$ neighbors of $\mathbf{x}^{(m)}$, namely,
$\mathbf{z}^{(m)}(u,v)$, $u=1,\cdots,d_t$, $v=1,\cdots,N$, are
computed as

\vspace{-3mm}
{\footnotesize
\begin{eqnarray}
\phi(\mathbf{z}^{(m)}(u,v)) &\hspace{-2.5mm}=& \hspace{-2.5mm} \big(\mathbf{x}^{(m)} + \mathbf{e}\big)^H \mathbf{R} \thinspace \big(\mathbf{x}^{(m)}+\mathbf{e}\big) - 2\Re\left(\big(\mathbf{x}^{(m)}+\mathbf{e}\big)^H \mathbf{y}_{MF}\right) \nonumber \\
& \hspace{-16mm} = & \hspace{-8mm} \phi(\mathbf{x}^{(m)})+2\Re\left( \mathbf{e}^H\mathbf{R}\thinspace\mathbf{x}^{(m)}\right) +\thinspace \mathbf{e}^H\mathbf{R}\thinspace \mathbf{e} - 2\Re\left(\mathbf{e}^H\mathbf{y}_{MF}\right) \nonumber \\
& \hspace{-16mm} = & \hspace{-8mm} \phi(\mathbf{x}^{(m)})+2\Re\left( \mathbf{e}^H\left(\mathbf{R}\thinspace\mathbf{x}^{(m)}-\mathbf{y}_{MF}\right)\right) +\thinspace \mathbf{e}^H\mathbf{R}\thinspace \mathbf{e} \nonumber \\
& \hspace{-16mm} = & \hspace{-8mm} \phi(\mathbf{x}^{(m)})+2\Re\left( \mathbf{e}^H \mathbf{f}^{(m)} \right) +\thinspace \mathbf{e}^H\mathbf{R}\thinspace \mathbf{e} \nonumber \\
& \hspace{-16mm} = & \hspace{-8mm} \phi(\mathbf{x}^{(m)})+ \underbrace{2 \Re\left( \thinspace e_u^*\thinspace {f}_u^{(m)}\right) + \thinspace \big|e_u\big|^2 \thinspace \mathbf{R}(u,u)}_{\Define \thinspace C(u,v)}\, ,
\label{cost}
\end{eqnarray}
}

\vspace{-7.0mm}
where  the last step follows since only one coordinate of $\mathbf{e}$
is non-zero, and $\mathbf{R}(u,u)$ is the $(u,u)$th element of
$\mathbf{R}$.  $\phi(\mathbf{x}^{(m)})$ on the RHS in (\ref{cost}) can
be dropped since it will not affect the cost minimization. Let
\begin{eqnarray}
(u_1,v_1)&=&
{\mbox{arg min}\atop{u,v}} \thinspace \thinspace \thinspace C(u,v).
\end{eqnarray}
The move $(u_1,v_1)$ is accepted if any one of the following two
conditions is satisfied:
\begin{eqnarray}
\label{mv1}
\phi(\mathbf{z}^{(m)}(u_1,v_1)) \,\,\,  < \,\,\, \phi(\mathbf{g}^{(m)}) \\
{\bf T}((u_1-1)M+q,v_1) \,\,\, = \,\,\, 0,
\label{move_accept}
\end{eqnarray}
where $q$ is such that $a_q = x_{u_1}^{(m)}, a_q \in \mathbb{A}$.
If move $(u_1,v_1)$ is not accepted (i.e., neither of the conditions in
(\ref{mv1}) and (\ref{move_accept}) is satisfied), find $(u_2,v_2)$ such
that
\begin{eqnarray}
(u_2,v_2)&=& {\mbox{arg min}\atop{u,v\thinspace \mbox{:} \thinspace u\ne u_1,v\ne v_1}} \thinspace \thinspace \thinspace C(u,v),
\end{eqnarray}
and check for acceptance of the $(u_2,v_2)$ move. If this also cannot be
accepted, repeat the procedure for $(u_3,v_3)$, and so on. If all the
$d_tN$ moves are tabu, then all the
$\textit{tabu\_matrix}$ entries are decremented by the minimum value in
the $\textit{tabu\_matrix}\,$; this goes on till one of the moves becomes
acceptable. Let $(u',v')$ be the index of the neighbor with the minimum
cost for which the move is permitted. Make
\begin{eqnarray}
\mathbf{x}^{(m+1)} & = & {\bf z}^{(m)}(u',v').
\end{eqnarray}
The variables $q',q'',v''$ are
implicitly defined by
{\small $a_{q'} = x_{u'}^{(m)} = w_{v''}(x_{u'}^{(m+1)})$},
and $a_{q''} = x_{u'}^{(m+1)}$, where $a_{q'}, a_{q''} \in \mathbb{A}$.
It is noted that in this {\em Step 1} of the algorithm, essentially
the best permissible vector-neighbor is chosen as the solution vector
for the next iteration.

\vspace{2.0mm}
{\em Step 2):}
The new solution vector obtained from {\em Step 1} is checked for repetition.
For the linear vector channel model in (\ref{eqn1}), repetition can be
checked by comparing the ML costs of the solutions in the previous
iterations. If there is a repetition, the length of the repetition from
the previous occurrence is found, the average length, $l_{rep}$, is updated,
and the tabu period $P$ is modified as $P=P+1$. If the number of iterations
elapsed since the last change of the value of $P$ exceeds $\beta l_{rep}$,
for a fixed $\beta > 0$, make $P=\max(1,P-1)$. After a move $(u',v')$ is
accepted, if $\phi( \mathbf{x}^{(m+1)})< \phi(\mathbf{g}^{(m)})$, make

\vspace{-3mm}
{\small
\begin{eqnarray}
{\bf T}((u'-1)M+q',v') \,\, = \,\,
{\bf T} ((u'-1)M+q'',v'') \,\, = \,\, 0,  \\
\label{tabumatrixupdate1}
\quad \mathbf{g}^{(m+1)} \,\, = \,\, \mathbf{x}^{(m+1)}, 
\end{eqnarray}
}

\vspace{-3mm}
else

\vspace{-3mm}
{\small
\begin{eqnarray}
{\bf T}((u'-1)M+q',v') \, = \,
{\bf T}((u'-1)M+q'',v'') \, = \, P+1,  \\
\label{tabumatrixupdate2}
\quad lflag = 1, \quad \mathbf{g}^{(m+1)} = \mathbf{g}^{(m)}\hspace{-1mm}.
\end{eqnarray}
}

\vspace{-3mm}
It is noted that this
{\em Step 2} of the algorithm implements the `reactive' part in the
search, by dynamically changing $P$.

\vspace{2mm}
{\em Step 3):}
Update the entries of the $\textit{tabu\_matrix}$ as
\begin{eqnarray}
\hspace{-6mm}
{\bf T}(r,s) & = & \max\{{\bf T}(r,s)-1,0\}, 
\end{eqnarray}
for {\small $r=1,\cdots,d_tM$}, {\small $s=1,\cdots,N,$}
and update $\mathbf{f}^{(m)}$ as
\begin{eqnarray}
\mathbf{f}^{(m+1)} & = & \mathbf{f}^{(m)}+\left(z_{u'}^{(m)}(u',v')-x_{u'}^{(m)}\right)\mathbf{R}_{u'},
\end{eqnarray}
where $\mathbf{R}_{u'}$ is the ${u'}$th column of $\mathbf{R}$.
The algorithm terminates in {\em Step 3} if the following stopping
criterion is satisfied, else it goes back to {\em Step 1}.

\vspace{2mm}
{\bfseries {\em Stopping criterion:}}
The algorithm can be stopped based on a fixed number of iterations.
Though convergence can be slow at low SNRs, it can be fast at moderate
to high SNRs. So rather than fixing a large number of iterations to stop
the algorithm irrespective of the SNR, we use an efficient stopping
criterion which makes use of the knowledge of the best ML cost found till
the current iteration, as follows. Since the ML criterion is to minimize
${\|\mathbf{Hx}-\mathbf{y}\|}^2$, the minimum value of the objective
function $\phi(\mathbf{x})$
is always greater than $-\mathbf{y}^H\mathbf{y}$. We stop the algorithm
when the least ML cost achieved in an iteration is within certain range
of the global minimum, which is $-\mathbf{y}^H\mathbf{y}$. We stop the
algorithm in the $m$th iteration, only if $lflag=1$ and the condition
\vspace{1mm}
\begin{eqnarray}
\frac{|\phi(\mathbf{g}^{(m)})-(-\mathbf{y}^H\mathbf{y})|}{|-\mathbf{y}^H\mathbf{y}|} \,\,\, < \,\,\, \alpha _{1}
\label{cons1}
\end{eqnarray}
is met with at least $\textit{min\_iter}$ iterations being completed to
make sure the search algorithm has `settled.' The bound is gradually
relaxed as the number of iterations increase and the algorithm is
terminated when
\begin{eqnarray}
\frac{|\phi(\mathbf{g}^{(m)})-(-\mathbf{y}^H\mathbf{y})|}{|-\mathbf{y}^H\mathbf{y}|} \,\,\, < \,\,\, m\alpha _2.
\label{cons2}
\end{eqnarray}
In (\ref{cons1}) and (\ref{cons2}), $\alpha_1$ and $\alpha_2$ are
positive constants.
In addition, we terminate the algorithm whenever the number of repetitions
of solutions exceeds $\textit{max\_rep}$. Also, the maximum number of
iterations is set to $\textit{max\_iter}$.

\subsection{RTS algorithm versus LAS algorithm}
It is noted that the likelihood ascent search (LAS) algorithm presented
in \cite{sun}-\cite{jstsp} is also a local neighborhood search based
algorithm, where the basic definition of neighborhood is the same as
in RTS. However, LAS differs from RTS in the following aspects:
$i)$ while the definition of neighborhood is static in LAS for all
iterations, in RTS, in addition to the basic neighborhood definition,
there is also a dynamic aspect to the neighborhood definition by way of
prohibiting certain vectors from being included in the neighbor list
(implemented through repetition checks/tabu period), and $ii)$ while
LAS gets trapped in the local minima that it first encounters and
declares this minima to be the final solution vector, RTS can potentially
find better minimas because of the escape strategy embedded in the algorithm
(by way of allowing to pick and move to the best neighbor even if that
neighbor has a lesser likelihood than the current solution vector).

It is further noted that a general version of LAS reported in \cite{jstsp},
termed as multistage LAS (MLAS), executes a different escape mechanism when
it encounters a local minima, by changing the neighborhood definition: it
considers vectors which differ in two or more coordinates (as opposed to
only one coordinate in the basic neighborhood definition) as neighbors.
On escaping from a local minima, the algorithm reverts back to the basic
neighborhood definition till the next local minima is encountered and
stops when no escape from a local minima is possible. Since the performance
gain of MLAS compared to LAS is found to be small, we limit our comparison
of RTS with only LAS. Our simulation results for the systems considered in
Sections \ref{sec3} to \ref{sec5} show that RTS performs better than LAS.

\section{RTS Performance in Large V-BLAST Systems }
\label{sec3}
Consider a V-BLAST MIMO system with $N_t$ transmit and $N_r$ receive antennas.
For this system, in the received signal model in (\ref{eqn1}), ${\bf x} \in
{\mathbb A}^{N_t}$ is the transmitted symbol vector, where ${\mathbb A}$
is the modulation alphabet, ${\bf H} \in {\mathbb C}^{N_r\times N_t}$
is the channel gain matrix whose entries are modeled as
$\mathcal C \mathcal N(0,1)$, ${\bf y} \in {\mathbb C}^{N_r}$ is the
received signal vector, and ${\bf n} \in {\mathbb C}^{N_r}$ is the
noise vector whose entries
are modeled as i.i.d $\mathcal C \mathcal N(0,\sigma^2=\frac{N_tE_s}{\gamma})$,
where $E_s$ is the average energy of the transmitted symbols and $\gamma$ is
the average received SNR per receive antenna.
We rewrite the complex system model in (\ref{eqn1}) as a real-valued
system as
\begin{eqnarray}
\label{SystemModelReal}
\tilde{\bf y} & = & \tilde{{\bf H}} \, \tilde{{\bf x}} + \tilde{{\bf n}},
\end{eqnarray}
where
\begin{eqnarray}
\label{SystemModelRealDef} \nonumber
\tilde{{\bf H}} = \left[\begin{array}{cc}\Re({\bf H}) \hspace{2mm} -\Im({\bf H}) \\
\Im({\bf H})  \hspace{5mm} \Re({\bf H}) \end{array}\right],
\quad
\tilde{{\bf y}} = \left[\begin{array}{c} \Re({\bf y}) \\ \Im({\bf y}) \end{array}\right], \\
\tilde{{\bf x}} = \left[\begin{array}{c} \Re({\bf x}) \\ \Im({\bf x}) \end{array}\right],
\quad
\tilde{{\bf n}} = \left[\begin{array}{c} \Re({\bf n}) \\ \Im({\bf n}) \end{array}\right].
\end{eqnarray}
We apply the RTS algorithm on the real-valued system model in
(\ref{SystemModelReal}) and estimate the transmitted symbol vector.
We note that the transmit and receive dimensions in the
linear vector channel in (\ref{SystemModelReal}) are $d_t=2N_t$ and
$d_r=2N_r$.

In this section, we present the uncoded BER performance of RTS
based detection of V-BLAST signals. Since the RTS algorithm is a heuristic,
analytical evaluation of the BER and convergence behavior is difficult.
So we evaluate the BER and convergence performance of the RTS algorithm
through simulations. The following RTS parameters are used in the
simulations for 4-QAM: MMSE initial vector, {\small $P_0=2, \beta=0.1,
\alpha_1=5\%, \alpha_2=0.05\%, \textit{max\_rep}=75, \textit{min\_iter}=20$.}
Perfect channel state information at the receiver (CSIR) and i.i.d.
fading are assumed.

\subsection{Convergence behavior of RTS in V-BLAST}
\label{conv}
\begin{figure}
\hspace{-6mm}
\includegraphics[width=3.95in, height=2.95in]{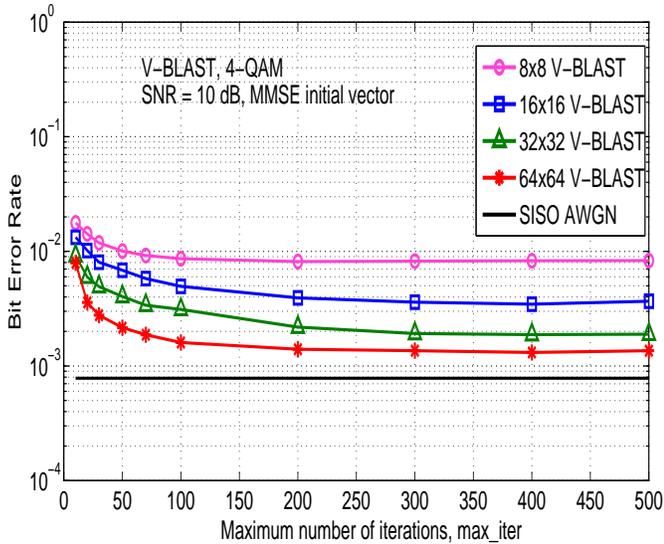}
\caption{
Uncoded BER performance of the RTS algorithm as a function of
maximum number of iterations, \textit{max\_iter}, in  $8\times 8,
16\times 16, 32\times 32$, and $64\times 64$ V-BLAST with 4-QAM at
SNR = 10 dB.
}
\vspace{-4mm}
\label{fig0x}
\end{figure}

In Fig. \ref{fig0x}, we plot the BER performance of the RTS algorithm
as a function of maximum number of iterations, \textit{max\_iter}, in
$8\times 8$, $16\times 16$, $32\times 32$, and $64\times 64$ V-BLAST
systems with 4-QAM at an average SNR of 10 dB.  Two main observations can
be made from Fig. \ref{fig0x}: $i)$ for the system parameters considered,
the BER converges (i.e., change in BER between successive iterations
becomes very small) for \textit{max\_iter} greater than 300, and $ii)$
the converged BER of RTS exhibits large-dimension behavior (i.e., converged
BER improves with increasing $N_t=N_r$); e.g., the converged BER improves
from $8.3\times 10^{-3}$ for $8\times 8$ V-BLAST to $1.3\times 10^{-3}$ for
$64\times 64$ V-BLAST. This improvement is quite significant considering
that the BER in SISO AWGN channel itself is $7.8\times 10^{-4}$ for 4-QAM.
We use \textit{max\_iter} to be 300 for 4-QAM in all the subsequent
simulations in this section.

\subsection{RTS versus LAS performance in V-BLAST}
\label{comp}
\begin{figure}
\hspace{-6mm}
\includegraphics[width=3.95in, height=2.95in]{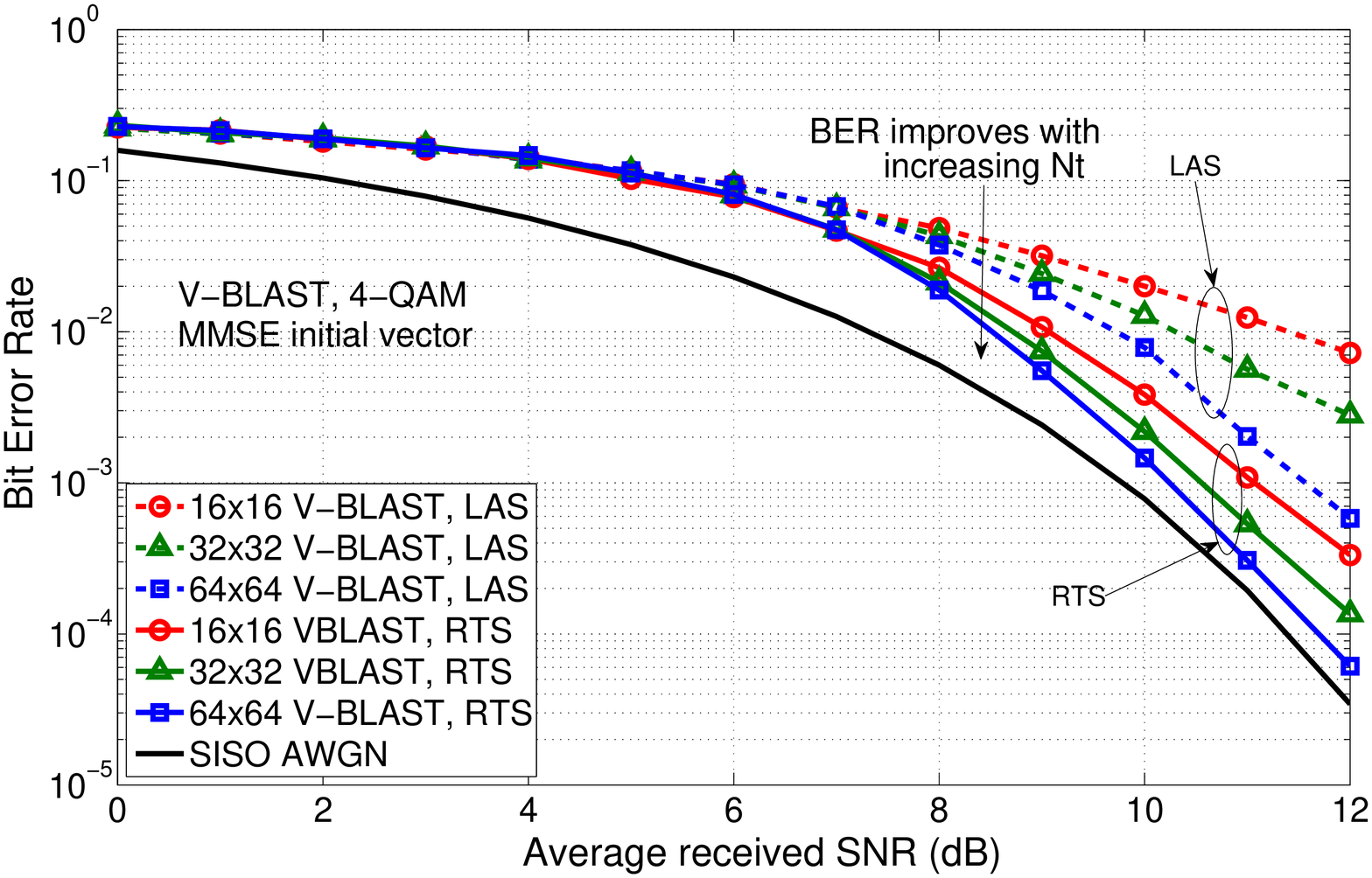}
\caption{
Uncoded BER performance of RTS detection of $16\times 16$,
$32\times 32$ and $64\times 64$ V-BLAST signals with 4-QAM.
}
\vspace{-4mm}
\label{fig2x}
\end{figure}

We next present the BER performance of the RTS algorithm in comparison
with that of the LAS algorithm presented in \cite{jstsp}. Figure \ref{fig2x}
shows the BER performance of RTS and LAS algorithms for $16\times 16$,
$32\times 32$ and $64\times 64$ V-BLAST with 4-QAM. It can be seen that
for the number of dimensions (i.e., $N_t$) considered, RTS performs better
than LAS; e.g., LAS requires 128 real dimensions (i.e., $64\times 64$
V-BLAST with 4-QAM) to achieve performance close to within 1.8 dB of SISO
AWGN performance at $10^{-3}$ BER, whereas RTS is able to achieve even
better closeness to SISO AWGN performance with just 32 real dimensions
(i.e., $16\times 16$ V-BLAST with 4-QAM). Also, in $64\times 64$ V-BLAST,
RTS achieves $10^{-3}$ BER at an SNR of just 0.4 dB away from SISO AWGN
performance. We note that RTS is able to achieve this better performance
because, while the bit/symbol-flipping strategies are similar in both RTS
and LAS, the inherent escape strategy in RTS allows it to move out of
local minimas and move towards better solutions. Consequently, RTS incurs
some extra complexity compared to LAS as detailed in the following
subsection.

\subsection{Complexity of RTS in V-BLAST}
\begin{figure}
\hspace{-6mm}
\includegraphics[width=3.95in, height=2.95in]{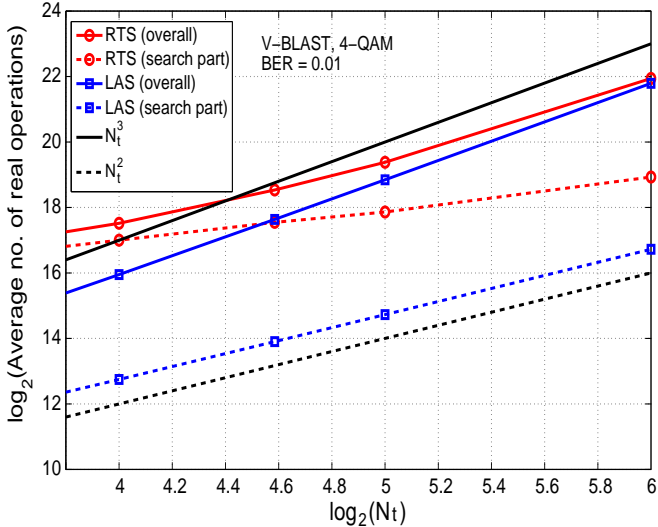}
\caption{
Complexity comparison of RTS and LAS algorithms in detection
of V-BLAST signals with 4-QAM at $10^{-2}$ BER.
}
\vspace{-4mm}
\label{fig2az}
\end{figure}

Here, we present the complexity of the RTS algorithm for detection in
V-BLAST. The total complexity comprises of three main components, namely,
$i)$ computation of the initial solution vector $\tilde{{\bf x}}^{(0)}$,
$ii)$ computation of $\tilde{{\bf H}}^T\tilde{{\bf H}}$, and $iii)$ the
reactive tabu search operation. The MMSE initial solution vector can be
computed in $O(N_t^2 N_r)$ complexity, i.e., in $O(N_tN_r)$ per-symbol
complexity since there are $N_t$ symbols per channel use. Likewise, the
computation of $\tilde{{\bf H}}^T\tilde{{\bf H}}$ can be done in
$O(N_tN_r)$ per-symbol complexity. We note that, since computation of
$\tilde{{\bf x}}^{(0)}$ and $\tilde{{\bf H}}^T\tilde{{\bf H}}$ are
needed in both RTS and LAS, the complexity components $i)$ and $ii)$
will be same for both these algorithms. We further note that, while the
complexity components $i)$ and $ii)$ are deterministic, the component
$iii)$, which is due to the search part alone, is random, and so we
obtained the average complexity of component $iii)$ through simulations.
Figure \ref{fig2az} shows the complexity plots for the
search part alone (i.e., component $iii)$) as well as the overall
complexity plots of the RTS and LAS algorithms for V-BLAST
with $N_t=N_r$ and 4-QAM at a BER of $10^{-2}$. From Fig. \ref{fig2az},
it can be observed that the RTS search part has a higher complexity
than the LAS search part. This is expected, because the RTS can escape
from a local minima and and look for better solutions, whereas LAS
settles in the first local minima itself.
However, it can be seen that since the overall complexity is dominated
by the computation of $\tilde{{\bf H}}^T \tilde{{\bf H}}$ and
$\tilde{{\bf x}}^{(0)}$, the difference in overall complexity between
RTS and LAS is not high.

\subsection{Comparison with variants of sphere decoders in V-BLAST}
\label{sec_new1}
\begin{figure}
\hspace{-6mm}
\includegraphics[width=3.95in, height=2.95in]{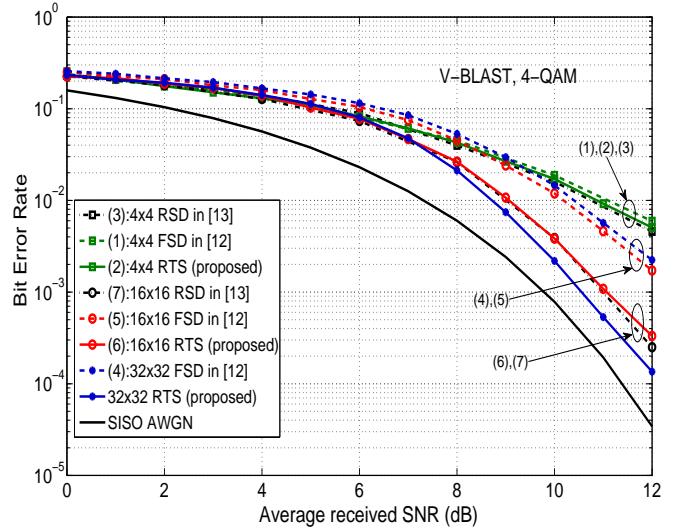}
\caption{
Comparison of uncoded BER performance of V-BLAST using RTS detection versus
fixed-complexity sphere decoding in \cite{fsd} and reduced complexity sphere
decoder in \cite{rsd} for $N_t=N_r=4,8,16,32$ and 4-QAM. RSD performance for
$32\times 32$ V-BLAST is not shown due to its high complexity.
}
\vspace{-4mm}
\label{fig2ax}
\end{figure}

In Fig. \ref{fig2ax}, we present a uncoded BER comparison of the RTS
detector with the fixed-complexity sphere decoder (FSD) presented in
\cite{fsd} for V-BLAST with $N_t=N_r=4,8,16,32$ and 4-QAM. The performance
of the reduced-complexity sphere decoder (RSD) presented in \cite{rsd} is
also plotted for $N_t=N_r=4,8,16$. We did not evaluate the performance of
RSD for $N_t=N_r=32$ due to its high complexity. Comparing the performances
of FSD, RSD and RTS in Fig. \ref{fig2ax}, we observe the following:
\begin{enumerate}
\item   Since the complexity of FSD is forced to be constant, the
        performance of FSD is compromised at low/medium SNRs compared to
        that of RSD (e.g., see plots for $N_t=N_r=16$, where RSD performs
        better than FSD by about 1 dB at $10^{-2}$ BER).
\item   Performance of RTS is very close to that of RSD (see plots of RSD
        and RTS for $N_t=N_r=16$). RTS achieves such good performance in
        large dimensions at a significantly lesser complexity compared to
        that of RSD (see complexity comparison in Table 1 for $16\times 16$
        V-BLAST).
\item   For large number of antennas (e.g., $N_t=N_r=32$), RSD complexity
        becomes prohibitively high, and so we do not show its performance
        for $32\times 32$ V-BLAST. However, we have shown the FSD and RTS
        performances for $32\times 32$ V-BLAST. It is seen that RTS performs
        significantly better than FSD (by about 1.5 dB at $10^{-2}$ BER);
        this is due to the sub-optimum nature of FSD that arises because
        of fixing its complexity, and due to the large-dimension behavior
        advantage of RTS.
        In addition, RTS achieves this better performance than FSD at a
        significantly lesser complexity compared to that of FSD (see details
        in the complexity comparison text in the following paragraphs and the
        $32\times 32$ system entries in Table 1).
\end{enumerate}

\vspace{1mm}
{\em {\bfseries Complexity comparison between RTS and FSD in V-BLAST:}}
The FSD algorithm in \cite{fsd} has two parts; an ordering part (similar
to that in V-BLAST algorithm) and a search part. The complexity of the
search part, which is random in conventional SD, is made constant in FSD
by fixing the number of search candidates irrespective of the SNR. The
ordering part has $O(N_t^3)$ complexity in $N_t$. Also, the algorithm
has $O(M^{\lceil \sqrt{N_t}-1\rceil})$ complexity in $M$ (i.e, alphabet
size) for $N_t=N_r$ \cite{fsd}. On the other hand, while RTS also has
$O(N_t^3)$ complexity in $N_t$ in a V-BLAST system, its complexity in $M$
is just $O(MN_t)$ since at most $(M-1)N_t$ neighbors need to be considered.
The exponential complexity of FSD in $\sqrt{N_t}$ makes it increasingly
prohibitive for increasing $N_t$. For e.g., for $N_t=N_r=32$ and 16-QAM, the
complexity of FSD, which is dominated by $O(M^{\lceil\sqrt{N_t}-1\rceil})$,
is $O(16^5)=O(2^{20})$. For the same system settings, the RTS complexity
is dominated by $O(N_t^3)$, which is $O(32^3) = O(2^{15})$. The differential
in complexity between RTS and FSD (in favor of RTS) widens further if 64-QAM
is considered.

A complexity comparison along with performance comparison between different
detectors is shown in Table 1, where we have presented the per-symbol
complexity (measured in number of real operations) and the SNR required
to achieve an uncoded BER of $10^{-2}$ in $4\times 4$, $8\times 8$,
$16\times 16$ and $32\times 32$ V-BLAST systems with 4-QAM. From Table 1,
we see that the complexity of FSD for $32\times 32$ V-BLAST is about an
order higher compared to that of RTS, due to the
$O(M^{\lceil\sqrt{N_t}-1\rceil})$ complexity of FSD. Also, even with this
higher complexity, FSD achieves poorer performance than RTS (i.e., FSD
needs about 1.5 dB more SNR than required by RTS to achieve $10^{-2}$ BER),
as described earlier.

\subsection{Higher-Order QAM Performance in V-BLAST}
\label{sec_new2}
\begin{figure}
\hspace{-6mm}
\includegraphics[width=3.95in, height=2.95in]{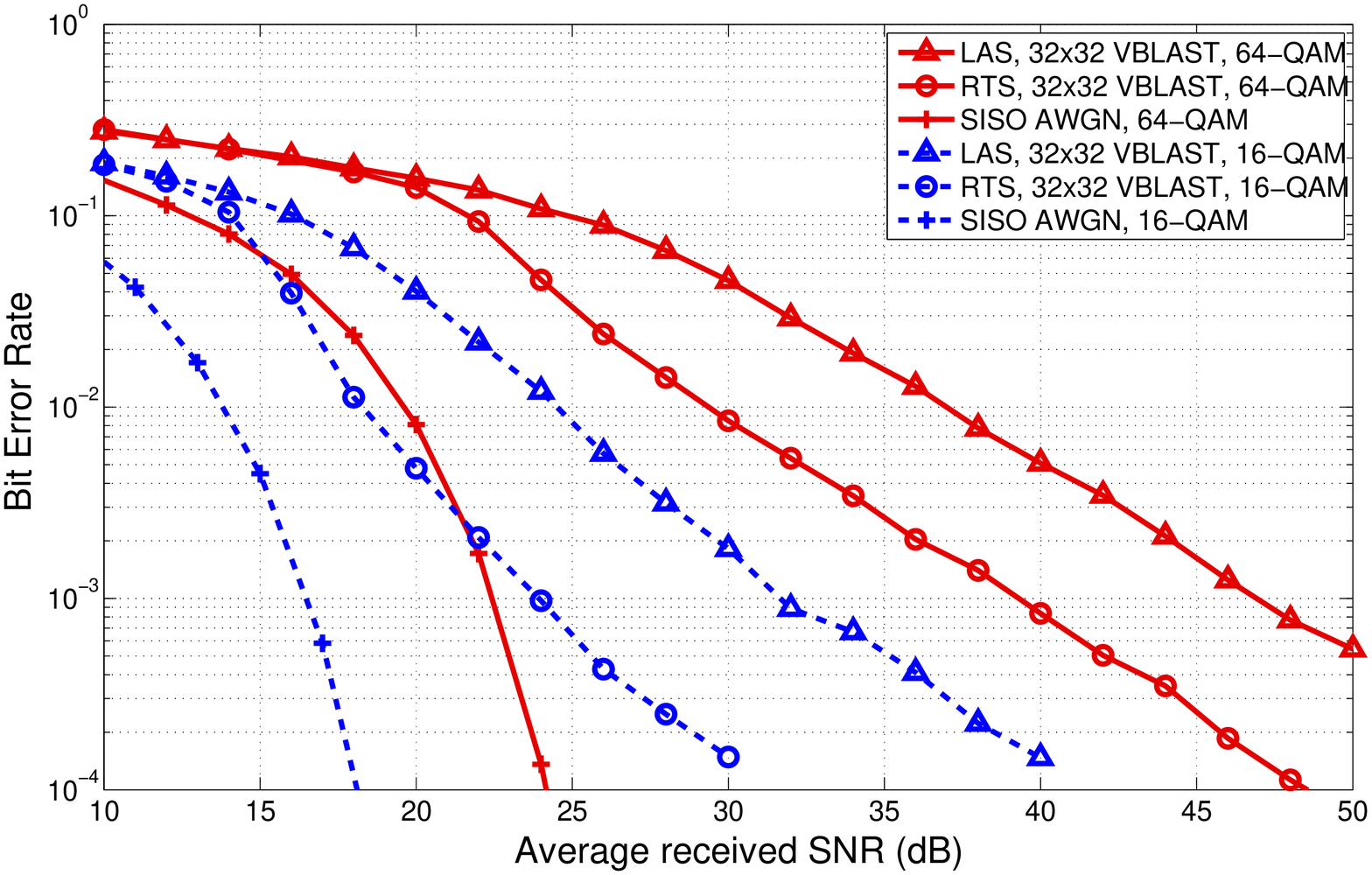}
\caption{
Uncoded BER performance of RTS and LAS algorithms for $32\times 32$ V-BLAST
system with 16-QAM and 64-QAM. FSD and RSD performances are not shown due
to their high complexities.
}
\vspace{-4mm}
\label{fig2ay}
\end{figure}

In Fig. \ref{fig2ay}, we illustrate the performance of RTS for
higher-order QAM in a $32\times 32$ V-BLAST system (16-QAM and 64-QAM
at spectral efficiencies of 128 bps/Hz and 192 bps/Hz).
We do not give the performance of FSD and RSD due to their high
complexities for the considered values of $N_t$ and $M$. As we
mentioned earlier, FSD complexity for $N_t=N_r=32$ and $M=64$ would
be $O(64^{\lceil\sqrt{32}-1\rceil}) = O(2^{30}$), which is prohibitive.
The complexities of RTS and LAS, on the other hand, scale well for such
large dimensions, allowing us to show their simulated BER performance
in Fig. \ref{fig2ay}. The following RTS parameters are used in the
simulations: MMSE initial vector, {\small $P_0=2, \beta=0.01$};
$({\small N=3, \alpha_1=0.3\%, \alpha_2=0.001\%, \textit{max\_rep}=250,
\textit{min\_iter}=}$\newline ${\small 30, \textit{max\_iter}=1000})$ 
for 16-QAM, and $({\small N = 2, \alpha_1=0.005\%}$, 
${\small \alpha_2=0.00005\%, \textit{max\_rep}=1000, \textit{min\_iter}=50, 
\textit{max\_iter}=}$ ${\small 3000})$ for 64-QAM.
The plots in Fig. \ref{fig2ay} show that
RTS performs better than LAS by about 6 dB at $10^{-2}$ BER for
16-QAM and 64-QAM.

\vspace{1cm}
\begin{table*}[t]
{\normalsize
\begin{center}
\begin{tabular}{|c||c|c||c|c||c|c||c|c||c|}
\hline
& \multicolumn{8}{|c||}{Per-symbol-complexity (PSC) in number of real operations and SNR} & \\ 
Detector & \multicolumn{8}{|c||}{required to achieve $10^{-2}$ uncoded BER for 4-QAM (Ref: Fig. \ref{fig2ax}) } & Order of \\\cline{2-9}
Algorithm & \multicolumn{2}{|c||}{4 $\times 4$ } & \multicolumn{2}{|c||}{8 $\times 8$} & \multicolumn{2}{|c||}{$16\times 16$}  & \multicolumn{2}{|c||}{$32\times 32$} & total complexity \\
\cline{2-9}
& PSC  & SNR & PSC & SNR & PSC & SNR & PSC & SNR & in $M$ and $N_t$ \\ \hline
&  &              &  & & & & & & \\
RTS & 5,540  & 10.9 dB & 9,469  & 9.7 dB & 11,730 & 9 dB & 21,320 & 8.8 dB  & {\footnotesize $O\big(MN_t\big) + O\big(N_t^3 \big)$} \\
(proposed) &  &             &  & &  & & & & \\ \hline
&  &              &  & & & & & &\\
FSD & 355  & 11 dB & 1,621  & 10.1 dB & 8,445 & 10.1 dB & 155,253 & 10.3 dB & {\footnotesize $O\big(M^{\lceil\sqrt{N_t}-1\rceil}\big) + O\big(N_t^3\big)$} \\
in \cite{fsd} &   &  &  & & & & & & \\ \hline
&  &              &  & & & & & & \\
RSD & 662  & 10.8 dB & 2,881 & 9.7 dB & 64,217 & 9 dB & - & - & -  \\
in \cite{rsd} &   &  &  & & & & & & \\ \hline
\end{tabular}
\label{tab2}
\vspace{2mm}
\caption{
Complexity and performance comparison of the RTS algorithm with the
FSD algorithm in \cite{fsd} and the RSD algorithm in \cite{rsd} in
$4\times 4$, $8\times 8$, $16\times 16$ and $32\times 32$ V-BLAST
with 4-QAM. RTS outperforms FSD in terms of complexity and performance
for large dimensions (e.g., $32\times 32$). For large $M$ and large
$N_t$, complexity of FSD gets prohibitively high. 
}
\label{table1}
\end{center}
}
\vspace{-8mm}
\end{table*}

\section{RTS Performance in Large Non-Orthogonal STBCs}
\label{sec4}
Large-MIMO systems that employ non-orthogonal STBCs from CDA
\cite{bsr},\cite{cda} are attractive because these STBCs can
simultaneously provide both {\em full rate} (i.e., $N_t$ complex
symbols per channel use, which is the same as in V-BLAST) as well
as {\em full transmit diversity} (V-BLAST does not provide transmit
diversity). The $2\times 2$ Golden code is a well known non-orthogonal
STBC from CDA for 2 transmit antennas \cite{gold05}. A non-orthogonal
STBC from CDA is a $N_t\times N_t$ matrix whose entries are formed
using linear combinations of various data symbols \cite{bsr}. Each STBC
matrix is constructed using $N_t^2$ data symbols, which are sent in
using $N_t$ transmit antennas in $N_t$ channel uses. The received
signal matrix can be vectorized and written in an equivalent real
system model of the form (\ref{SystemModelReal}), where the number of
transmit and receive dimensions are $d_t=2N_t^2$ and $d_r=2N_tN_r$,
respectively, for QAM \cite{jstsp}.

\vspace{1mm}
High spectral efficiencies can be achieved using {\em large} non-orthogonal
STBCs from CDA. For e.g., a $16\times 16$ STBC from CDA has 256 complex
symbols in it with {\em 512 real dimensions}; with 16-QAM and rate-3/4
turbo code, this system offers a high spectral efficiency of 48 bps/Hz.
Variants of sphere decoding (e.g., FSD \cite{fsd} and RSD \cite{rsd})
do not scale well to decode signals with hundreds of
dimensions\footnote{Since FSD and RSD complexities are prohibitive
to decode signals with hundreds of dimensions, we do not present the
performance of FSD and RSD for large non-orthogonal STBCs.}. In
\cite{jstsp}, we have shown that the LAS algorithm can scale well to
such hundreds of dimensions while achieving good performance. In this
section, we show that RTS also scales well in complexity in decoding
large non-orthogonal STBCs from CDA having hundreds of dimensions, while
achieving even better performance than LAS.

{\em RTS complexity in decoding non-orthogonal STBCs from CDA:}
Here again, 
$\tilde{{\bf H}}^T\tilde{{\bf H}}$ computation complexity dominates the
overall complexity compared to the search complexity. Note that there
$2N_t^2$ transmit and $2N_tN_r$ receive dimensions, and $N_t^2$ symbols
per STBC. Exploiting the permutation nature of the weight matrices of
the non-orthogonal STBCs from CDA \cite{jstsp}, the per-symbol complexity
of computing $\tilde{{\bf H}}^T\tilde{{\bf H}}$, and hence the overall
per-symbol complexity in RTS decoding of non-orthogonal STBCs from CDA
is $O(N_t^2N_r)$.

\vspace{1mm}
In the following subsections, we present the BER performance of RTS in
decoding non-orthogonal STBCs. The following parameters are used in the
simulations for 4-QAM: MMSE initial vector, {\small $P_0=2, \beta=1,
\alpha_1=5\%, \alpha_2=0.05\%, \textit{max\_rep}=75, \textit{min\_iter}=20,
\textit{max\_iter}=300$.}

\subsection{RTS versus LAS performance in decoding non-orthogonal STBCs}
\begin{figure}
\hspace{-6mm}
\includegraphics[width=3.95in, height=2.95in]{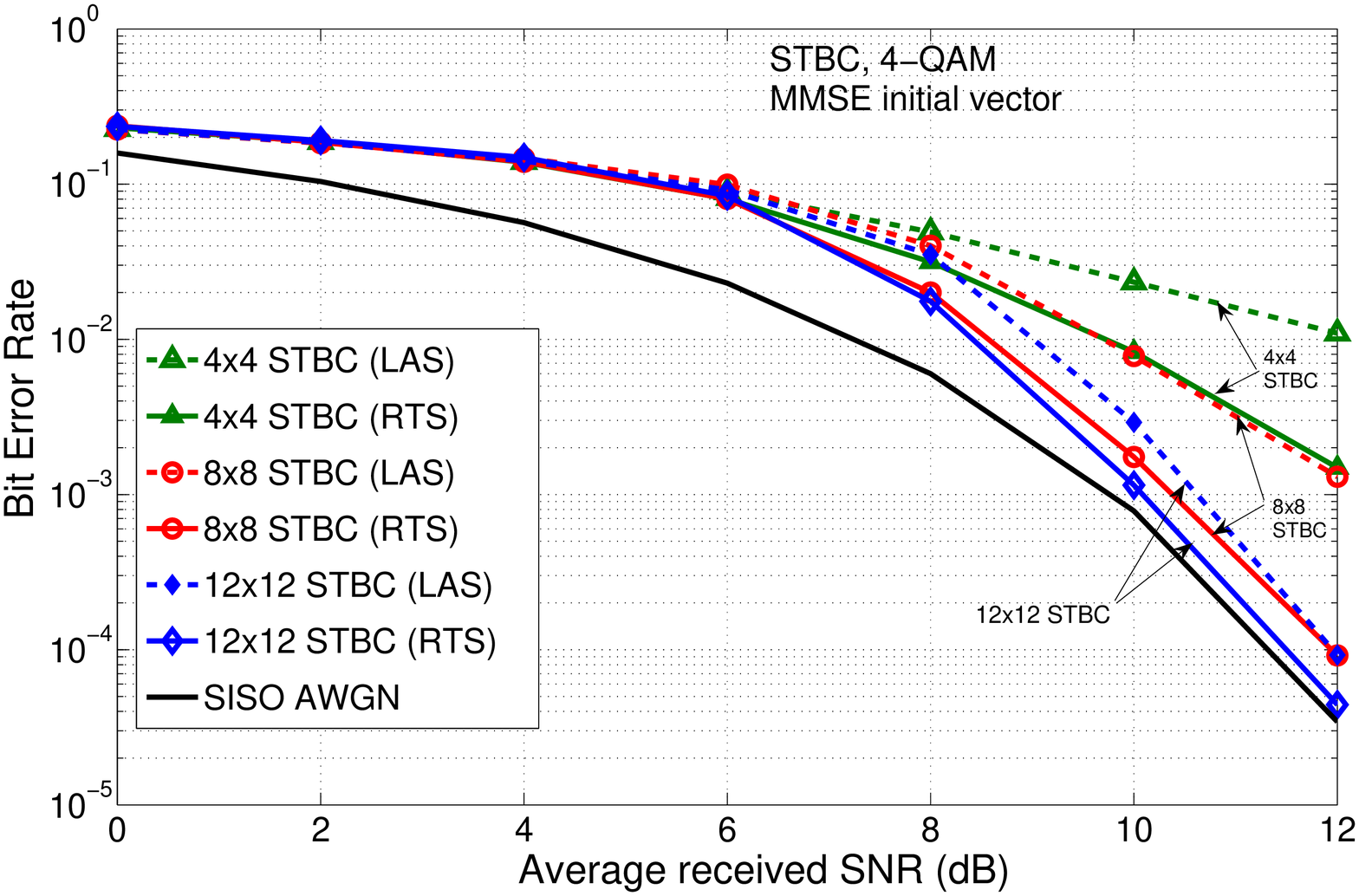}
\caption{
Uncoded BER of RTS decoding of $4\times 4$, $8\times 8$,
and $12\times 12$ non-orthogonal STBCs from CDA for $N_t=N_r$ and
4-QAM.
}
\vspace{-4mm}
\label{fig1}
\end{figure}

In Fig. \ref{fig1}, we plot the uncoded BER of the RTS algorithm
as a function of average received SNR in decoding $4\times 4$
(32 dimensions), $8\times 8$ (128 dimensions) and $12\times 12$
(288 dimensions) non-orthogonal STBCs from CDA for 4-QAM and $N_t=N_r$.
Perfect CSIR and i.i.d fading are assumed. For the same settings,
performance of the LAS algorithm is also plotted for comparison.
MMSE initial vector is used in both RTS and LAS. As a lower bound on
performance, we have plotted the BER performance on a SISO AWGN channel
as well. From Fig. \ref{fig1}, it can be observed that the BER of RTS
improves and approaches SISO AWGN performance as
$N_t\hspace{-0.5mm}=\hspace{-0.5mm}N_r$ (i.e., {\small STBC} size) is
increased; e.g., with $12\times 12$ STBC having 288 dimensions, RTS
decoding is able to achieve close to within 0.4 dB from SISO AWGN
performance at $10^{-3}$ uncoded BER. Also, as in the case of V-BLAST,
RTS is found to perform better than LAS in decoding non-orthogonal
STBCs as well. In the case of 16-QAM also, RTS performs better than LAS
as can be seen in Fig. \ref{16qam}, where the following parameters are
used in the simulations: MMSE initial vector, {\small $P_0=2,
\beta=1, N=3, \alpha_1=0.1\%, \alpha_2=0.002\%, \textit{max\_rep}=75,
\textit{min\_iter}=30, \textit{max\_iter}=800$.}

\subsection{Turbo coded BER performance of RTS}
Figure \ref{fig3} shows the rate-3/4 turbo coded BER performance of RTS
decoding of $12\times 12$ non-orthogonal STBC from CDA
with $N_t=N_r$ and 4-QAM
(corresponding to a spectral efficiency of 18 bps/Hz), under perfect CSIR
and i.i.d fading. The theoretical minimum SNR required to achieve 18 bps/Hz
spectral efficiency on a
$N_t\hspace{-0.25mm}=\hspace{-0.25mm}N_r\hspace{-0.25mm}=\hspace{-0.25mm}12$
MIMO channel with perfect CSIR and i.i.d fading is 4.27 dB (obtained
through simulation of the ergodic MIMO capacity formula \cite{jafarkhani}).
From Fig. \ref{fig3}, it is seen that RTS decoding is able to achieve
vertical fall in coded BER close to within about 5 dB from the theoretical
minimum SNR, which is a good nearness to capacity performance. This nearness
to capacity can be further improved by 1 to 1.5 dB if soft decision values,
proposed in \cite{jstsp}, are fed to the turbo decoder. Also, the performance
of RTS is about 1 dB better than that of LAS at $2\times 10^{-4}$ coded BER
for the same system settings.

\subsection{Iterative RTS Decoding/Channel Estimation}
\begin{figure}
\hspace{-6mm}
\includegraphics[width=3.95in, height=2.95in]{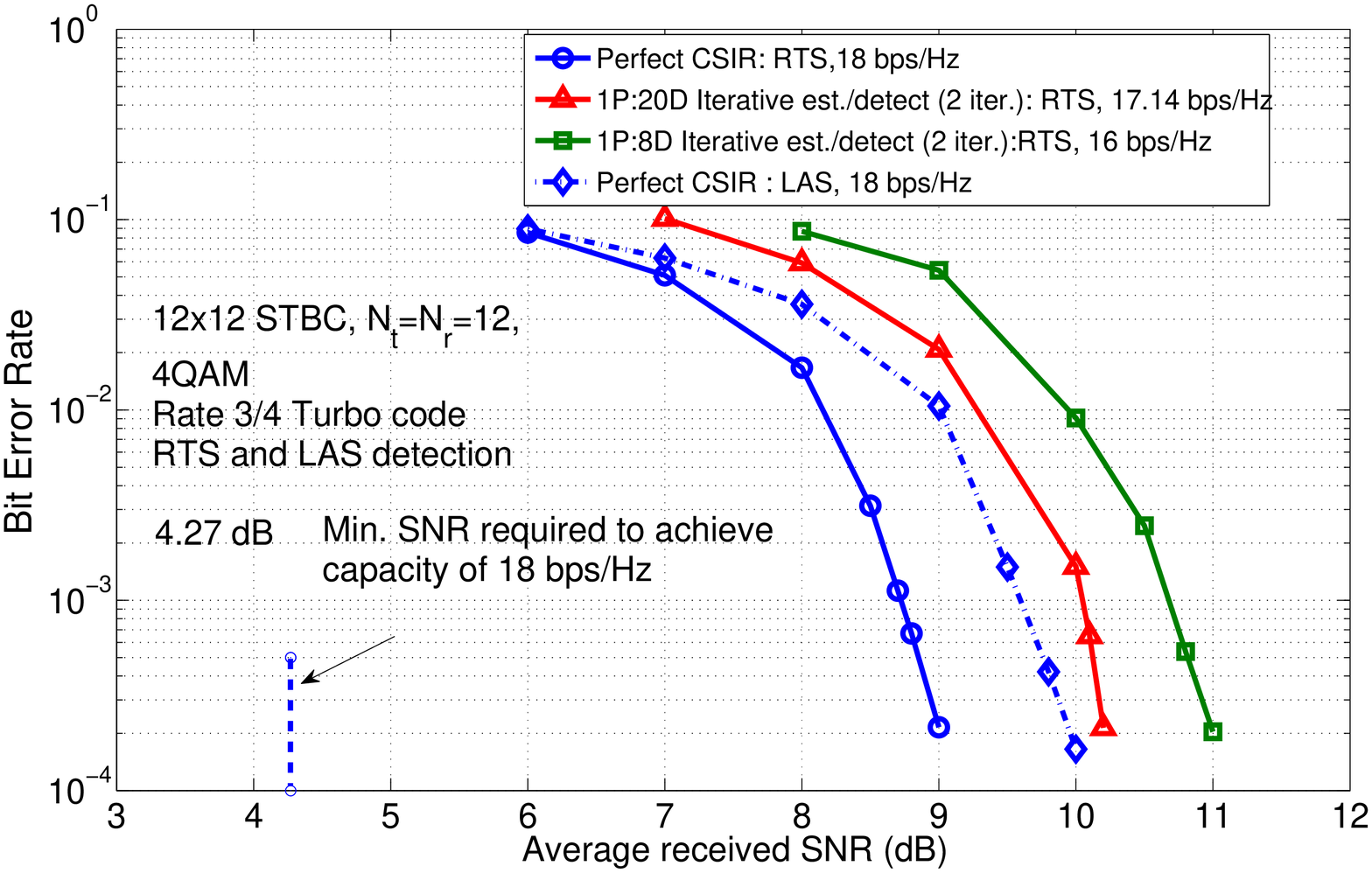}
\caption{
Turbo coded BER of RTS decoding of $12\times 12$ non-orthogonal
STBC from CDA with $N_t=N_r$, 4-QAM, rate-3/4 turbo code, and 18 bps/Hz
with perfect CSIR and estimated CSIR.
}
\vspace{-4mm}
\label{fig3}
\end{figure}

Next, we relax the perfect CSIR assumption by considering a training
based iterative RTS decoding/channel estimation scheme. Transmission is
carried out in frames, where one {\small $N_t\times N_t$} pilot matrix
(for training purposes) followed by $N_d$ data STBC matrices are sent
in each frame \cite{jstsp}. One frame length, $T$, (taken to be the
channel coherence time) is $T=(N_d+1)N_t$ channel uses. The proposed
scheme works as follows: $i)$ obtain an MMSE estimate of the channel
matrix during the pilot phase, $ii)$ use the estimated channel matrix
to decode the data STBC matrices using RTS, $iii)$ use the decoded STBCs
to estimate the channel matrix again, and $iv)$ iterate between channel
estimation and RTS decoding for a certain number of times. For $12\times 12$
STBC from CDA, in addition to perfect CSIR performance, Fig. \ref{fig3} also
shows the performance with CSIR estimated using the above iterative RTS
decoding/channel estimation scheme for $N_d=8$ and $N_d=20$. 2 iterations
between RTS decoding and channel estimation are used. With $N_d=20$ (which
corresponds to large coherence times, i.e., slow fading) the BER
and bps/Hz with estimated CSIR get closer to those with perfect CSIR.

\subsection{Effect of MIMO Spatial Correlation}
\begin{figure}
\hspace{-6mm}
\includegraphics[width=3.95in, height=2.95in]{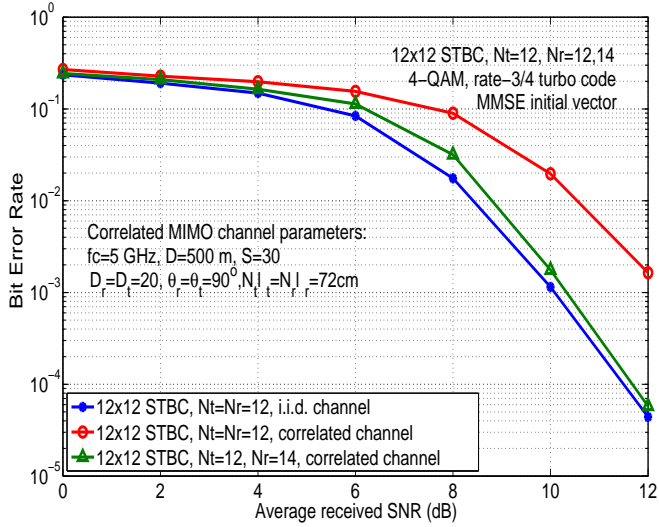}
\caption{
Effect of spatial correlation on the performance of RTS decoding
of $12\times 12$ STBC from CDA with $N_t=12$, $N_r=12,14$, 4-QAM, rate-3/4
turbo code, 18 bps/Hz.  $f_c=5$ GHz, $D=500$ m, $S=30$, $D_t=D_r=20$ m,
$\theta_t=\theta_r=90^\circ$, $N_rl_r=N_tl_t=72$ cm.
}
\vspace{-4mm}
\label{fig4}
\end{figure}

In all the previous performance and complexity plots, we assumed i.i.d
fading. But spatial
correlation at transmit/receive antennas and the structure of scattering
and propagation environment can affect the rank structure of the MIMO
channel resulting in degraded performance \cite{mimo1},\cite{mimo2}. We
relaxed the i.i.d. fading assumption by considering the correlated MIMO
channel model proposed by Gesbert et al in \cite{mimo2}, which takes into
account carrier frequency ($f_c$), spacing between antenna elements {\small
($l_t,l_r$)}, distance between transmit and receive antennas ($D$), and
scattering environment. In Fig. \ref{fig4}, we plot the uncoded BER of
RTS decoding of $12\times 12$ STBC from CDA with perfect CSIR in $i)$
i.i.d. fading, and $ii)$ correlated MIMO fading model in \cite{mimo2}.
It is seen that, compared to i.i.d fading, there is a performance loss
in spatial correlation for $N_t=N_r=12$; further, use of more receive
antennas ($N_r=14, N_t=12$) alleviates this loss in performance.

\section{RTS Equalizer for MIMO-ISI Channels}
\label{sec5}
In this section, we consider the adoption and performance of the RTS
algorithm in another communication scheme, where large dimensions
are created in time due to the highly frequency selective nature of
the channel, i.e., large number (tens to hundreds) of multipath
components (MPC), as can typically happen in UWB channels
\cite{uwb0},\cite{uwb3}.

\begin{figure}
\vspace{6mm}
\includegraphics[width=3.5in, height=1.80in]{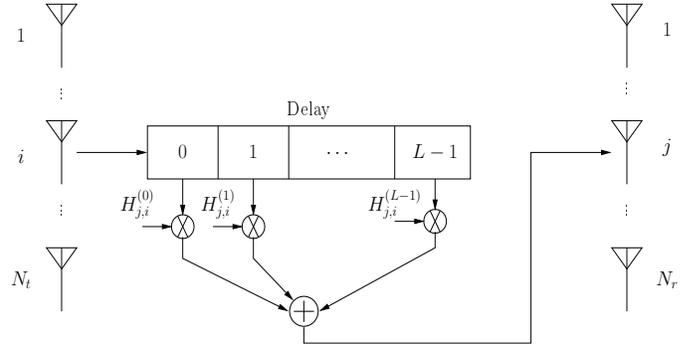}
\caption{MIMO-ISI channel model.
}
\label{fig_uwb}
\end{figure}

Consider a frequency-selective MIMO channel with $N_t$ transmit and
$N_r$ receive antennas (Fig. \ref{fig_uwb}). 
Let $L$ denote the number of MPCs. Data is
transmitted in frames, where each frame has $K$ data symbols
preceded by a cyclic prefix (CP) of length $L$ symbols, $K\geq L$.
While CP avoids inter-frame interference, there will be ISI within
the frame. Let ${\bf x}_q \in {\mathbb A}^{N_t}$ be the transmitted
symbol at time $q$, $0 \leq q \leq K-1$, where ${\mathbb A}$ is the
transmit symbol alphabet, which is taken to be $M$-QAM. The received
signal vector at time $q$ can be written as
\begin{eqnarray}
\label{eq1}
{\mathbf y}_q &=& \sum\limits_{l=0}^{L-1}{\mathbf H}_l{\mathbf x}_{q-l}+{\mathbf w}_q, \quad \quad
q = 0,\cdots,K-1,
\end{eqnarray}
where ${\mathbf y}_q \in \mathbb{C}^{N_r\times 1}$,
${\mathbf H}_l \in \mathbb{C}^{N_r \times N_t}$ is the channel gain matrix
for the $l$th MPC. The entries of ${\mathbf H}_l$ are assumed to be random
with distribution $\mathcal{CN}(0,1)$. It is further assumed that
${\mathbf H}_l$, $l=0,\cdots,L-1$ do not change for one frame duration.
${\mathbf w}_q \in \mathbb{C}^{N_r\times 1}$ is the additive white Gaussian
noise vector at time $q$, whose entries are independent, each with variance
$N_0$. The CP will render the linearly convolving channel to a circularly
convolving one, and so the channel will be multiplicative in frequency
domain. Because of the CP, the received signal in frequency domain, for
the $i$th frequency index ($0 \leq i \leq K-1$), can be written as
\begin{eqnarray}
\label{eq3}
{\mathbf r}_i & = & {\mathbf G}_i \ {\mathbf u}_i + {\mathbf v}_i,
\end{eqnarray}
where
${\mathbf r}_i = \frac{1}{\sqrt{K}} \sum\limits_{q=0}^{K-1} e^{\frac{-2\pi {\mathfrak j}qi}{K}} {\mathbf y}_q, \, \,$
${\mathfrak j}=\sqrt{-1}, \, \,$
${\mathbf u}_i = \frac{1}{\sqrt{K}} \sum\limits_{q=0}^{K-1} e^{\frac{-2\pi {\mathfrak j}qi}{K}} {\mathbf x}_q, \, \,$
${\mathbf v}_i = \frac{1}{\sqrt{K}} \sum\limits_{q=0}^{K-1} e^{\frac{-2\pi {\mathfrak j}qi}{K}} {\mathbf w}_q, \, \,$
and
${\mathbf G}_i=\sum\limits_{l=0}^{L-1}e^{\frac{-2\pi{\mathfrak j}li}{K}} {\mathbf H}_l$.
Stacking the $K$ vectors ${\mathbf r}_i$, $i=0,\cdots,K-1$, we can write
\begin{eqnarray}
\label{eqChannelModel}
{\mathbf r} & = & \underbrace{{\mathbf{GF}}}_{\Define \,\, {\mathbf H}_{eff}}{\mathbf x}_{eff} + {\mathbf v}_{eff},
\end{eqnarray}
where
{\normalsize
\[{\mathbf r}=\left[ \begin{array}{c} {\mathbf r}_0 \\ {\mathbf r}_1\\
\vdots \\ {\mathbf r}_{{\small K-1}} \end{array} \hspace{-2mm}\right]
\mathrm{,} \quad
{\mathbf G}=\left[ \begin{array}{cc} \begin{array}{ll} {\mathbf G}_0  & \\ & {\mathbf G}_1 \end{array} & {\mathbf 0} \\
{\mathbf 0}  &  \begin{array}{ll} \ddots & \\  & {\mathbf G}_{{\small K-1}} \end{array} \end{array}\hspace{-3mm} \right]\mathrm{,} \nonumber \]
}
{\normalsize
\[
{\mathbf x}_{eff}=\left[ \begin{array}{c} {\mathbf x}_0 \\ {\mathbf x}_1 \\ \vdots \\ {\mathbf x}_{{\small K-1}} \end{array} \hspace{-2mm}\right]
\mathrm{,} \quad
{\mathbf v}_{eff}=\left[ \begin{array}{c}{\mathbf v}_0 \\ {\mathbf v}_1\\ \vdots \\ {\mathbf v}_{{\small K-1}} \end{array}\right],
\]
\begin{eqnarray*}
{\mathbf F} & \hspace{-3.5mm} = & \hspace{-3.5mm}\frac{1}{\sqrt{K}}\left[ \begin{array}{llll} \rho_{{\small 0,0}}{\mathbf I}_{N_t} & \rho_{{\small 1,0}}{\mathbf I}_{N_t} & \cdots & \rho_{{\small K-1,0}}{\mathbf I}_{N_t} \\ \rho_{{\small 0,1}}{\mathbf I}_{N_t} & \rho_{{\small 1,2}}{\mathbf I}_{N_t} & \cdots & \rho_{{\small K-1,1}}{\mathbf I}_{N_t} \\
\vdots  & \vdots  & \cdots & \vdots   \\
\rho_{{\small 0,K-1}}{\mathbf I}_{N_t} & \rho_{{\small 1,K-1}}{\mathbf I}_{N_t} & \cdots & \rho_{{\small K-1,K-1}}{\mathbf I}_{N_t} \end{array} \hspace{-2mm}\right] \nonumber \\
& = & \frac{1}{\sqrt{K}}{\mathbf D}_K \otimes {\mathbf I}_{N_t},
\end{eqnarray*}
}

\vspace{-2mm}
where $\rho_{q,i}=e^{\frac{-2\pi{\mathfrak j}qi}{K}}$, ${\mathbf D}_K$ is
the $K$-point DFT matrix and $\otimes$ denotes the Kronecker product.
The received signal model in (\ref{eqChannelModel}) can be rewritten in
real form with $d_t=2N_tK$ and $d_r=2N_rK$. RTS algorithm is applied on
this real-valued system model.

{\em Initial vector using FD-MMSE equalizer:} The detected symbol vector
obtained using frequency domain (FD) MMSE equalization can be used as the
initial vector to the RTS algorithm. The FD-MMSE equalizer on the $i$th
frequency employs MMSE nulling as
\begin{eqnarray}
\label{eq4}
{\widehat{\mathbf u}}_i & = & \Big({\mathbf G}_i^H {\mathbf G}_i+
\frac{N_0}{E_s}{\mathbf I}_{N_t} \Big)^{-1}{\mathbf G}_i^H{\mathbf r}_i,
\quad 0\leq i \leq K-1,
\label{fdmmse1}
\end{eqnarray}
where $E_s$ is the average
energy of a transmitted symbol. The ${\widehat{\mathbf u}}_i$'s are
transformed back to time domain using $K$-point IDFT to obtain
an estimate of the transmitted symbol vector as
\begin{eqnarray}
\label{eq5}
{\widehat{\mathbf x}}_q & = & \frac{1}{\sqrt{K}}\sum_{i=0}^{K-1} e^{\frac{2\pi {\mathfrak j}qi}{K}} {\widehat{\mathbf u}}_i, \quad 0\leq q \leq K-1,
\label{fdmmse2}
\end{eqnarray}
which are used to form the initial vector to the RTS algorithm.

\subsection{Performance Results and Discussions}
We evaluated the BER performance of the proposed RTS equalizer in a
$4\times 4$ MIMO V-BLAST system with 4-QAM as a function of average
$E_b/N_0$ per receive antenna, through simulations. We have assumed
uniform power delay profile (i.e., all the $L$ paths are assumed to be
of equal energy). We evaluated the performance for various number of
delay paths, $L$, and frame sizes, $K$, keeping $L/K$ constant. It is
noted that the system becomes a `large-dimension system' when $L$ and
$K$ are increased keeping $L/K$ fixed. The FD-MMSE equalizer output is
used as the initial vector for both RTS and LAS. The following RTS
parameters are used: $P_0=2; \beta=1; \alpha_1=0.03; max\_rep=75;
min\_iter=30$.  For $K=64$ and 128, $max\_iter=300$ and
$\alpha_2=0.00075$. For $K=512$, $max\_iter=500$ and $\alpha_2=0.0004$.

\begin{figure}
\hspace{-6mm}
\includegraphics[width=3.95in, height=2.95in]{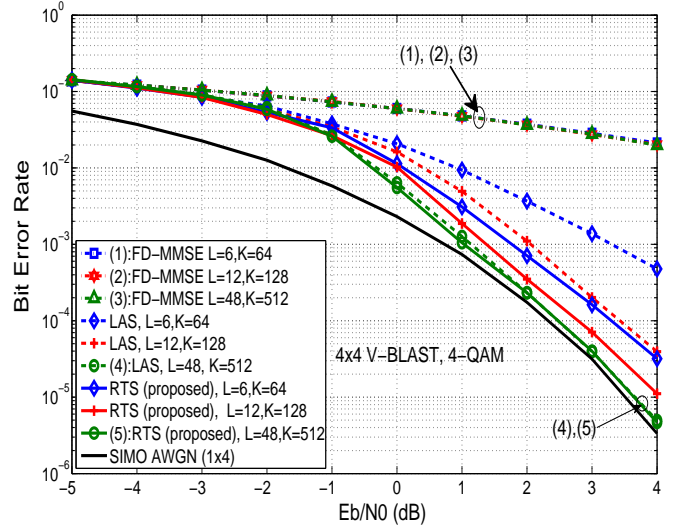}
\caption{
Comparison of the BER performance of the proposed RTS equalizer
with those of LAS equalizer and FD-MMSE equalizer in a $4\times 4$
V-BLAST system with 4-QAM for different number of MPCs ($L$) and frame
sizes ($K$), keeping $L/K$ constant. Uniform power delay profile.
}
\vspace{-4mm}
\label{fig6}
\end{figure}

In Fig. \ref{fig6}, we plot the uncoded BER of the RTS equalizer for
$(L=6, K=64)$, $(L=12,K=128)$, and $(L=48, K=$ $512)$, $L/K=0.09375$.
Note that for $(L=48, K=$ $512)$, the number of transmit dimensions
is $d_t=2N_tK=2\times 4\times 512 = 4096$ dimensions. Since FSD and
RSD complexities are prohibitive for number of dimensions in the
thousands, we do not give their performances. In addition to the
performance of RTS, we have given the performance of $i)$ the FD-MMSE
equalizer (without any subsequent search), $ii)$ LAS equalizer, and
$iii)$ single-input multiple-output (SIMO) AWGN with $N_r=4$ (which
can be viewed as a good lower bound on the best detector performance).
It is seen that the performance of the FD-MMSE equalizer is poor.
However, the subsequent search operations carried out in RTS and LAS
result in significantly improved performance for increasing $L$, $K$.
Both RTS and LAS show large-dimension behavior in this system also
(i.e., BER improves for increasing $L$, $K$, keeping $L/K$ fixed).
For a given $L$, RTS performs better than LAS. For e.g., at $10^{-3}$
BER, RTS performs better by about 1.5 dB and 0.8 dB compared LAS for
($L=6, K=64)$ and ($L=12,K=128$),  respectively.
We note that the per-symbol complexity of FD-MMSE (i.e.,
initial vector) computation is $O(KN_t+N_t^2)$. The per-symbol
complexity of $\tilde{{\bf H}}^T\tilde{{\bf H}}$ computation is
$O(K^2N_t)$. The per-symbol search complexities for RTS, obtained
by simulations, is $O(KN_t)$. So the overall per-symbol complexity
of the RTS equalizer is $O(K^2N_t) + O(KN_t+N_t^2)$.

\section{Conclusions}
\label{sec6}
We conclude by highlighting some recent trends in high spectral efficiency
MIMO systems/measurements with large number of antennas to bring out the
contextual importance and relevance of the work presented in this paper.
1) NTT DoCoMo has already field demonstrated a $12\times 12$ V-BLAST system
operating at 5 Gbps data rate and 50 bps/Hz spectral efficiency in 4.6 GHz
band at a mobile speed of 10 Km/hr \cite{docomo}.
2) Evolution of WiFi standards (evolution from IEEE 802.11n to IEEE 802.11ac
to achieve multi-gigabit rate transmissions in 5 GHz band) now considers
$16\times 16$ MIMO operation; e.g., see $16\times 16$ MIMO indoor channel
sounding measurements at 5.17 GHz reported in \cite{dot11ac} for
consideration in WiFi standards. 3) $64 \times 64$ MIMO channel sounding
measurements at 5 GHz in indoor environments have been reported in \cite{hut}.
We note that, while the RF/antenna technologies/measurements for large-MIMO
systems are getting matured, there is lack of current focus on development
of low-complexity baseband algorithms for detection and channel estimation
for large-MIMO systems (MIMO systems with 16 or more antennas) to
reap their high spectral efficiency benefits. A vast body of MIMO detection
literature is heavily focused on $4\times 4$ (in some cases $8\times 8$)
MIMO. Algorithms suited for large-MIMO signal detection and their
performance have started appearing in the literature recently (e.g.,
\cite{jsac},\cite{jstsp}). Here, we showed that the RTS algorithm presented
in this paper achieves even better performance than the LAS algorithm
presented in \cite{jstsp} (e.g., 6 dB better performance in $32\times 32$
V-BLAST with 16- and 64-QAM in Fig.  \ref{fig2ay}). We also showed that the
considered sphere decoding variants (FSD, RSD) either performed poorly and/or
did not scale well for large-dimension detection (e.g., see $32\times 32$
V-BLAST plots and complexities in Fig. \ref{fig2ax} and Table 1). The
large-dimension behavior of the RTS algorithm has other potential
applications, like the low-complexity equalization in severely delay-spread
UWB systems (with thousands of dimensions) presented in this paper. Finally,
we note that algorithms for low-complexity, high-performance
large-dimen\-sion signal processing for communication applications is a
promising research direction.

\end{document}